\documentclass[useAMS,usenatbib,usegraphicx]{mn2e}
\usepackage{color}
\def\bmath#1{\mbox{\boldmath$#1$}}
\def\revise#1{{#1}}
\def\aap{A\&A}
\def\mnras{MNRAS}
\def\apj{ApJ}
\def\apjl{ApJ}
\def\nat{Nature}
\def\pasj{PASJ}
\def\sun{\odot}

\title[An origin of arc structures]{An origin of arc structures deeply embedded in dense molecular cloud cores}

\author[T. Matsumoto et al.]{Tomoaki Matsumoto$^{1}$\thanks{E-mail: matsu@hosei.ac.jp} 
  Toshikazu Onishi$^{2}$,
  Kazuki Tokuda$^{2}$, and
  Shu-ichiro Inutsuka$^{3}$\\
$^{1}$Faculty of Humanity and Environment, Hosei University, Fujimi, Chiyoda-ku, Tokyo 102-8160, Japan\\
$^{2}$Department of Physical Science, Graduate School of Science, Osaka Prefecture University, 1-1 Gakuen-cho, Naka-ku, Sakai, Osaka 599-8531, Japan\\
$^{3}$Department of Physics, Nagoya University, Chikusa-ku, Nagoya 464-8602, Japan
}

\begin{document}


\pagerange{\pageref{firstpage}--\pageref{lastpage}} \pubyear{2015}

\maketitle

\label{firstpage}

\begin{abstract}
We investigated the formation of arc-like structures in the infalling envelope
around protostars, motivated by the recent Atacama Large Millimeter/Submillimeter Array (ALMA) observations of
the high-density molecular cloud core, MC27/L1521F.  We performed
self-gravitational hydrodynamical numerical simulations with an
adaptive mesh refinement code.  A filamentary cloud with a 0.1~pc
width fragments into cloud cores because of perturbations due to weak
turbulence.  The cloud core undergoes gravitational collapse to form
multiple protostars, and gravitational torque from the orbiting
protostars produces arc structures extending up to a 1000~AU scale.
As well as on a spatial extent, the velocity ranges of the arc
structures, $\sim0.5\,\mathrm{km\,s}^{-1}$, are
in agreement with the ALMA observations.  
We also found that circumstellar disks are often misaligned in
triple system. The misalignment is caused by the tidal interaction
between the protostars when they undergo close encounters because of a
highly eccentric orbit of the tight binary pair. 
\end{abstract}

\begin{keywords}
hydrodynamics -- ISM: clouds -- ISM: kinematics and
  dynamics -- stars: formation -- turbulence.
\end{keywords}

\section{Introduction}

It has been thought that multiple stars are formed through
fragmentation from a high density portion of a molecular cloud core.
The fragmentation occurs deep inside the molecular cloud core
and it has not ever been observed directly.

MC27 \citep{Mizuno94,Onishi99,Onishi02} or L1521F \citep{Codella97} is
a dense cloud core in Taurus and observations have suggested that
MC27/L1521F is in a very early stage of star formation \citep{Onishi99}.
Recently \citet{Tokuda14} performed Atacama Large Millimeter/Submillimeter Array (ALMA)
 observations on dust
continuum emission and molecular rotational lines toward MC27/L1521F.
One of their new findings is an arc structure with a $\sim
  2000~\mathrm{AU}$ length at the center of the molecular cloud core,
suggesting that the arc structure can be attributed for a dynamical
interaction between envelope gas and dense gas condensations. They
also suggested that the turbulence promotes fragmentation of cloud
cores during the collapse as shown by the numerical simulations of
turbulent cloud cores
\citep[e.g.,][]{Bate02,Goodwin04}.  In these
numerical simulations, the orbital motion of fragments excite spiral arms
in the center of the cloud core.  However, these studies focus mainly
on the physical fragmentation process, while the observable features, such as
the velocity structure of spiral arms, are still poorly understood.

In this paper, the protostellar collapse of a turbulent core is
calculated from a parent filamentary cloud
by using adaptive mesh refinement (AMR) simulations.
We reproduced arc structures on a scale of about 1000 AU at the central part of
the molecular cloud core.  Based on these simulations, 
the formation mechanism of the arc structures is discussed. 
This paper is organized as follows.
In section~\ref{sec:method_and_models}, the
models and simulation methods are presented.  The results of
the simulations are shown in section~\ref{sec:results}, and they are discussed
in section~\ref{sec:discussion}.

\section{Models and methods}
\label{sec:method_and_models}

A filamentary molecular cloud is considered as the initial condition for this study
because recent observations have revealed that filaments
are basic compositions of molecular clouds
\citep[e.g.,][]{Andre10,Hacar13}.  The filament is
assumed to be infinitely long and in an equilibrium state where
thermal pressure supports an isothermal cloud against its
self-gravity.  The cloud therefore has a density distribution of $\rho(R) =
\rho_0 (1+ R^2/R_0^2)^{-2}$, where $R$, $\rho_0$ and $R_0$ are a
cylindrical radius, density on the filamentary axis, and a scale
height of the filament, respectively
\citep{Stodolkiewicz63,Ostriker64}.  We set $R_c =
0.05\,\mathrm{pc}$, which mimics filaments observed by the 
Herschel survey \citep{Arzoumanian11}.  The gas temperature
was assumed to be $T = 10$~K at the initial stage and the corresponding
sound speed is $c_s = 0.190\,\mathrm{km}\,\mathrm{s}^{-1}$.  The
density on the filamentary axis is given by $\rho_0 = 2c_s^2/(\pi G
R_0^2) = 1.45\times 10^{-19}\,\mathrm{g}\,\mathrm{cm}^{-3}$
(the corresponding number density is $n_0 =
3.79\times10^4\,\mathrm{cm}^{-3}$).  The barotropic equation of state
was assumed to be $P(\rho) = c_s^2 \rho + \kappa \rho ^{7/5}$ with $\kappa
= c_s^2 \rho_\mathrm{cr}^{-2/5}$, where the critical density is set at
$\rho_\mathrm{cr} = 10^{-13}\,\mathrm{g}\,\mathrm{cm}^{-3}$
(the corresponding number density is $n_\mathrm{cr} = 2.62\times
10^{10}\,\mathrm{cm}^{-3}$), which was taken from the numerical results
of \citet{Masunaga98}.  The magnetic field was ignored for simplicity.
The computational domain is a cubic box with side lengths $L=4\lambda_\mathrm{max}
=1.56~\mathrm{pc}$, where $\lambda_\mathrm{max} (=7.81 R_0)$ is 
the wavelength of the most unstable perturbation against 
fragmentation of the filament \citep{Nagasawa87}.
Periodic boundary conditions were imposed.

Turbulence was imposed at the initial stage.
The initial
velocity field is incompressible with a power spectrum of $P(k)
\propto k^{-4}$, generated according to \citet{Dubinski95}, where $k$
is the wavenumber.  This power spectrum results in a velocity
dispersion of $\sigma(\lambda) \propto \lambda^{1/2}$, in agreement
with the Larson scaling relations \citep{Larson81}.  
The root mean square (rms) Mach number in the computational domain
was set at unity.
Note that the rms Mach number on a 0.1~pc scale is expected to be
$\sim 0.25$ according to the scaling relations, and 
the turbulence is therefore subsonic on the filament scale.
Such a subsonic
turbulence was suggested by the narrow molecular line widths of dense
cores in Taurus \citep{Onishi98}.

The evolution of the filamentary cloud was calculated using
three-dimensional AMR code, SFUMATO \citep{Matsumoto07}.  The
hydrodynamic scheme was modified to have a third order of accuracy in
space and a second order in time. 
The computational domain is resolved on a base grid of $l =0$ with 
$256^3$ cells.  The maximum grid level was set at $l = 11$.
The cell width is $\Delta x_\mathrm{min} = 0.613$~AU on the finest grid of $l =
11$, compared with $\Delta x_\mathrm{max} = 1.26\times10^3$~AU on the base grid of $l =0$.
The Jeans condition was employed as a refinement criterion;
blocks are refined when the Jeans length is shorter than 8 times the cell width, i.e.,
$\lambda_J < 8 \Delta x $, where $\lambda_J$ is the Jeans length
 \citep[c.f.,][]{Truelove97}.

The sink particle is introduced as a sub-grid model of protostars. 
The detailed implementations of the sink particles are shown in \citet{Matsumoto15}.
The critical density for sink particle formation is set at 
\revise{
$\rho_\mathrm{sink} = 1\times 10^{-11}\,\mathrm{g}\,\mathrm{cm}^{-3}$
($n_\mathrm{sink} = 2.62\times 10^{12}\,\mathrm{cm}^{-3}$), }
and the sink radius is set at
$r_\mathrm{sink} = 4 \Delta x_\mathrm{min} = 2.45~\mathrm{AU}$.

\section{Results}
\label{sec:results}

\begin{figure*}
\begin{center}
\includegraphics[scale=0.6]{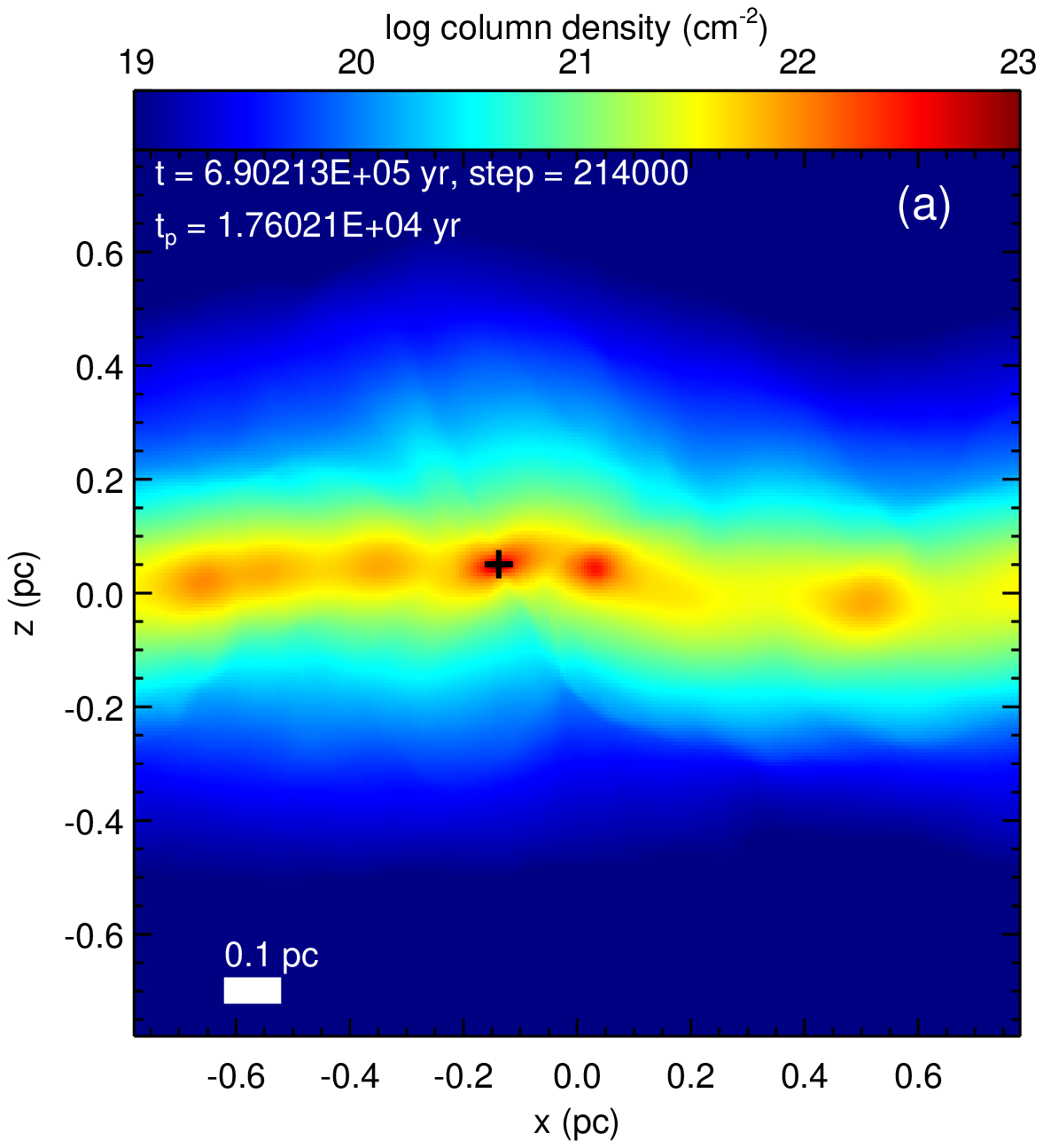}
\includegraphics[scale=0.6]{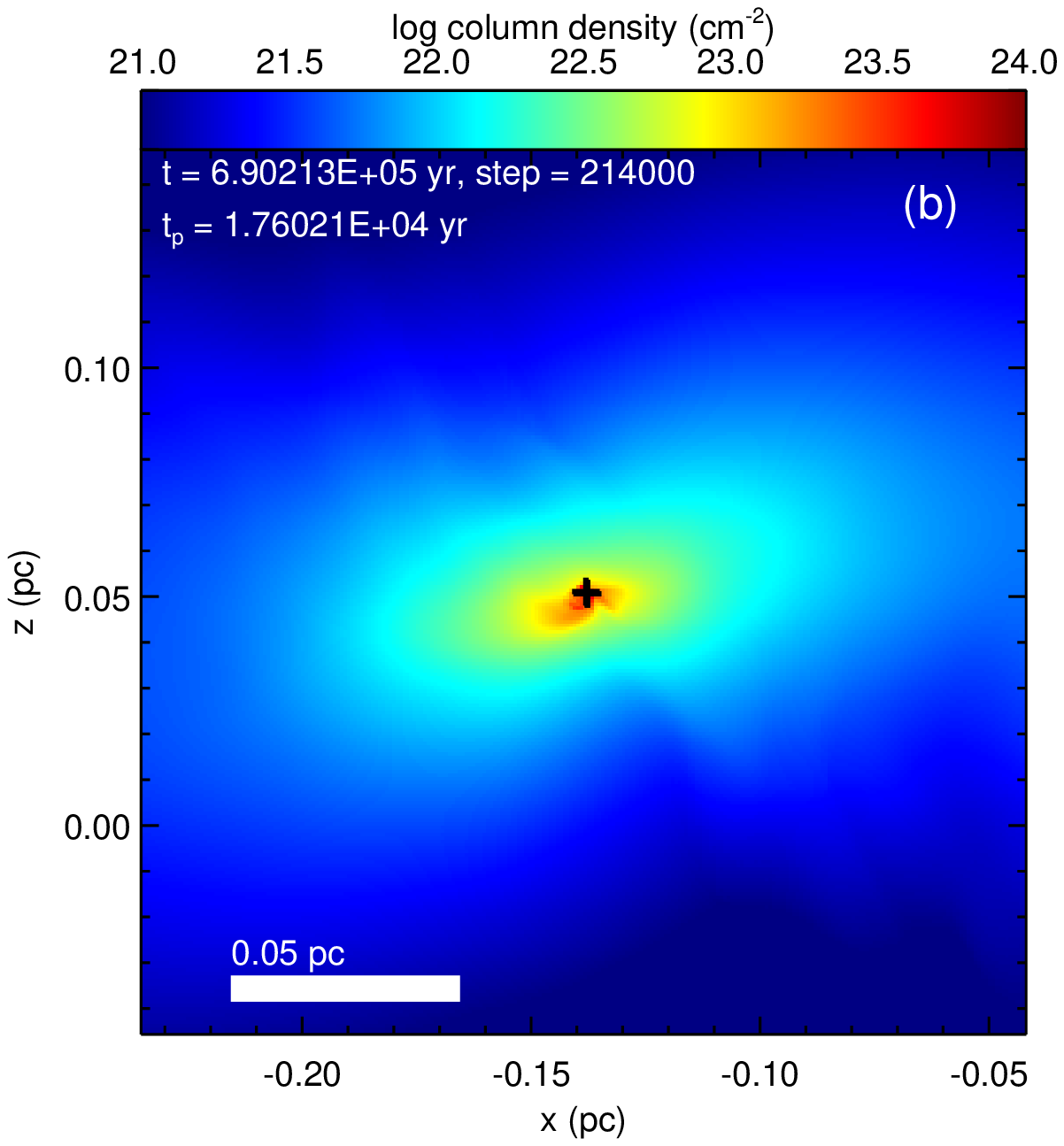}\\
\includegraphics[scale=0.6]{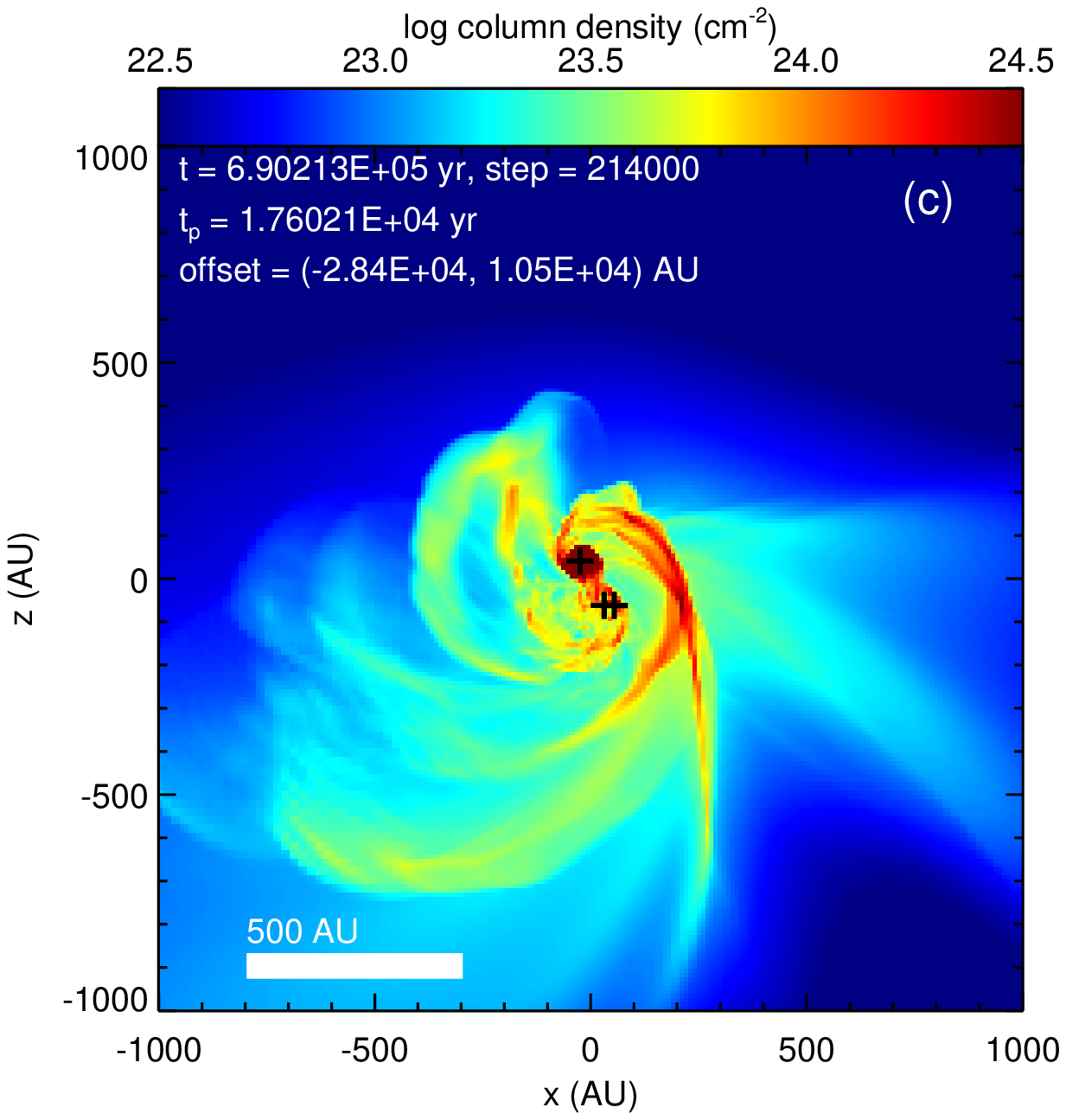}
\includegraphics[scale=0.6]{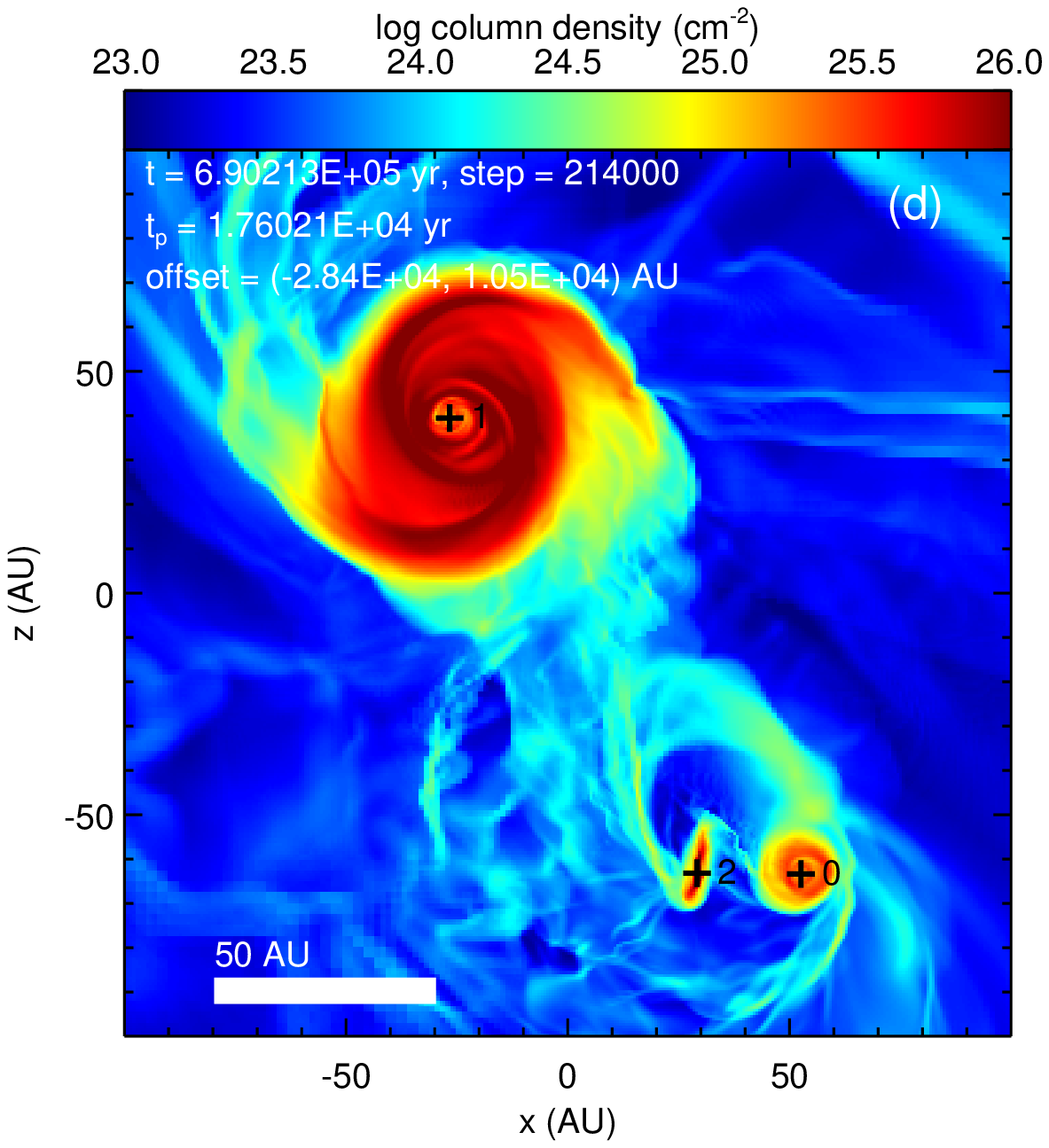}
\end{center}
\caption{ 
Column density distribution along the $y$-direction on 
four different spatial scales at $t = 0.690$~Myr ($t_p = 1.760 \times 10^4$~yr).  
The color scales depict the column densities.  The black crosses indicate the positions of
the sink particles. In panels (c) and (d), the coordinates
are offset so that the center of each panel coincides with the center
of mass of the sink particles. 
The sink particles are labeled with identification numbers 0, 1, and
2 in panel (d).
\label{columndensity_plot.eps}
}
\end{figure*}

Figure~\ref{columndensity_plot.eps} shows the column density distribution
along the $y$-direction for four different spatial scales at
$t = 6.90\times 10^5$~yr, or $t_p = 1.760 \times 10^4$~yr, where $t_p$
denotes an elapsed time after the first sink particle formation.
Figure~\ref{columndensity_plot.eps}(a) shows the whole computational
domain.  The filamentary cloud fragments into the several cloud
cores, one of which forms stars.  The filamentary cloud is disturbed by
the turbulence, which was imposed on the initial condition.
Figure~\ref{columndensity_plot.eps}(b) shows the cloud core.
The cloud core has a sharply peaked density
distribution in agreement with molecular line observations toward
MC27/L1521F \citep{Onishi99}.
The sink particles are associated with the peak position. 
The cloud core has a three-dimensional prolated shape and 
it exhibits smooth isodensity surfaces, in agreement with \citet{Matsumoto11}.

\begin{figure}
\includegraphics[scale=0.6]{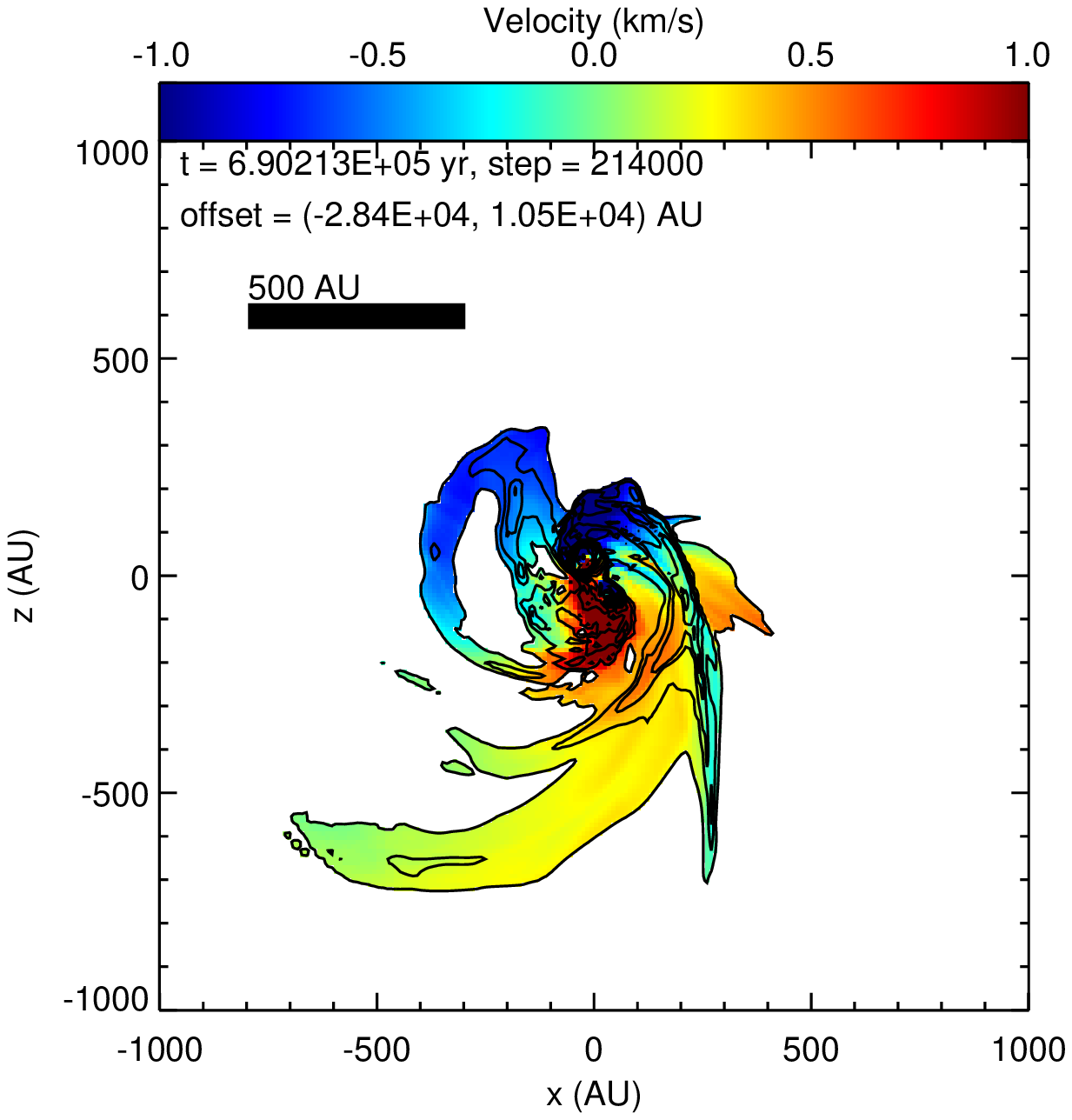}
\caption{ 
Line-of-sight velocity distribution in the arc structures.
The color scale shows the velocity along the line of sight.
The contour shows the column density on a logarithmic scale with 
$\log (N/\mathrm{cm}^{-2}) = 23.4, 23.6, \cdots$, where $N$ denotes
the column density.
The velocity is shown only in the regions where
the column density is greater than the lowest contour level.
\label{los_velocity_plot.eps}
}
\end{figure}

On a smaller scale as shown in Figure~\ref{columndensity_plot.eps}(c), a triple
system of sink particles can be seen.  The sink particles are
associated with several spiral arms, which correspond to the
arc structures on a 1000~AU scale.  The spiral arms are caused by
the gravitational interaction between the sink particles and the envelope
gas.  The gas is accelerated by the gravitational torque from the
orbiting sink particles, and the gas has a supersonic
velocity of typically $\sim 0.5\,\mathrm{km\,s}^{-1}$, exciting shock waves of
arc-shapes.
Figure~\ref{los_velocity_plot.eps} shows the distribution of 
the line-of-sight velocity, which was evaluated as a first moment of
velocity,
$\int v_y (\bmath{r}) \rho(\bmath{r})dy/\int \rho(\bmath{r}) dy$,
assuming that the gas is optically thin.
This figure is comparable to Figure~4 from \citet{Tokuda14}.
The arc shape and the velocity range are in agreement with the
observations of \citet{Tokuda14}.

\revise{ 
Note that the disk-like structure on a 1000~AU scale, which is shown as
a region with $\log N \, \mathrm{[cm^{-2}]}\ga 23.5$ in
Figure~\ref{columndensity_plot.eps}(c), is an infalling envelope.  
Its structure is not supported by rotation, and the 
rotation velocity is comparable to the infall velocity there.
The $Q$-value of \citet{Toomore64} is larger than unity, indicating that
the infalling envelope as well as the arcs are not significantly self-gravitating.
}

Figure~\ref{columndensity_plot.eps}(d) is a close-up view of the
triple system.  The sink particles are labeled by identification
numbers 0, 1 and 2 in order of formation epoch (hereafter referred to as Sinks~0, 1, and 2).
Sinks~0 and 2 constitute a close binary with a separation of
$20-30$~AU, while Sink~1 orbits around the close binary with
a long separation of $100-200$~AU.
At the stage as shown in Figure~\ref{columndensity_plot.eps}, Sinks~0, 1, and 2 have masses of
$0.20\,M_\odot$, $0.32\,M_\odot$, and $0.13\,M_\odot$, respectively. 
Sink~1 has the largest mass among the sink particles,
and it has the largest circumstellar disk with a
radius of $\sim 30\,\mathrm{AU}$,
\revise{
while the other sink particles have small circumstellar disks with radii of
$\sim 7\,\mathrm{AU}$.
  The circumstellar disk of Sink~1 is relatively massive;
  the mass ratio between the disk and the sink particle remains at 
  $M_\mathrm{disk}/M_\mathrm{star} \sim 0.2$ for a long period of $t_p
  \ga 700\,\mathrm{yr}$, while other sink particles take ratios of 
  $M_\mathrm{disk}/M_\mathrm{star} \sim 0.003 - 0.01$.
  The massive disk of Sink~1 has the spiral arms caused by the gravitational instability; the disk exhibits
  the $Q$-value less than unity.
  They promote a high accretion 
  rate onto Sink~1, $\sim 2\times10^{-5} M_\sun \mathrm{yr}^{-1}$.
  }

The circumstellar disk of Sink~2 is inclined with respect to
the other circumstellar disks.  
The misalignment of the circumstellar disk is caused by tidal
interactions from Sink~0 when Sink~2 undergoes sccessive close
encounters with Sink~0 \revise{\citep[c.f.,][]{Heller93}}.
These close encounters are caused by the highly eccentric orbit of the tight
binary pair of Sinks~0 and 2.  
\revise{During the evolution, rapid changes in the disk orientation 
occur several times for the disks of Sinks~0 and 2.}

\begin{figure}
\includegraphics[scale=0.5]{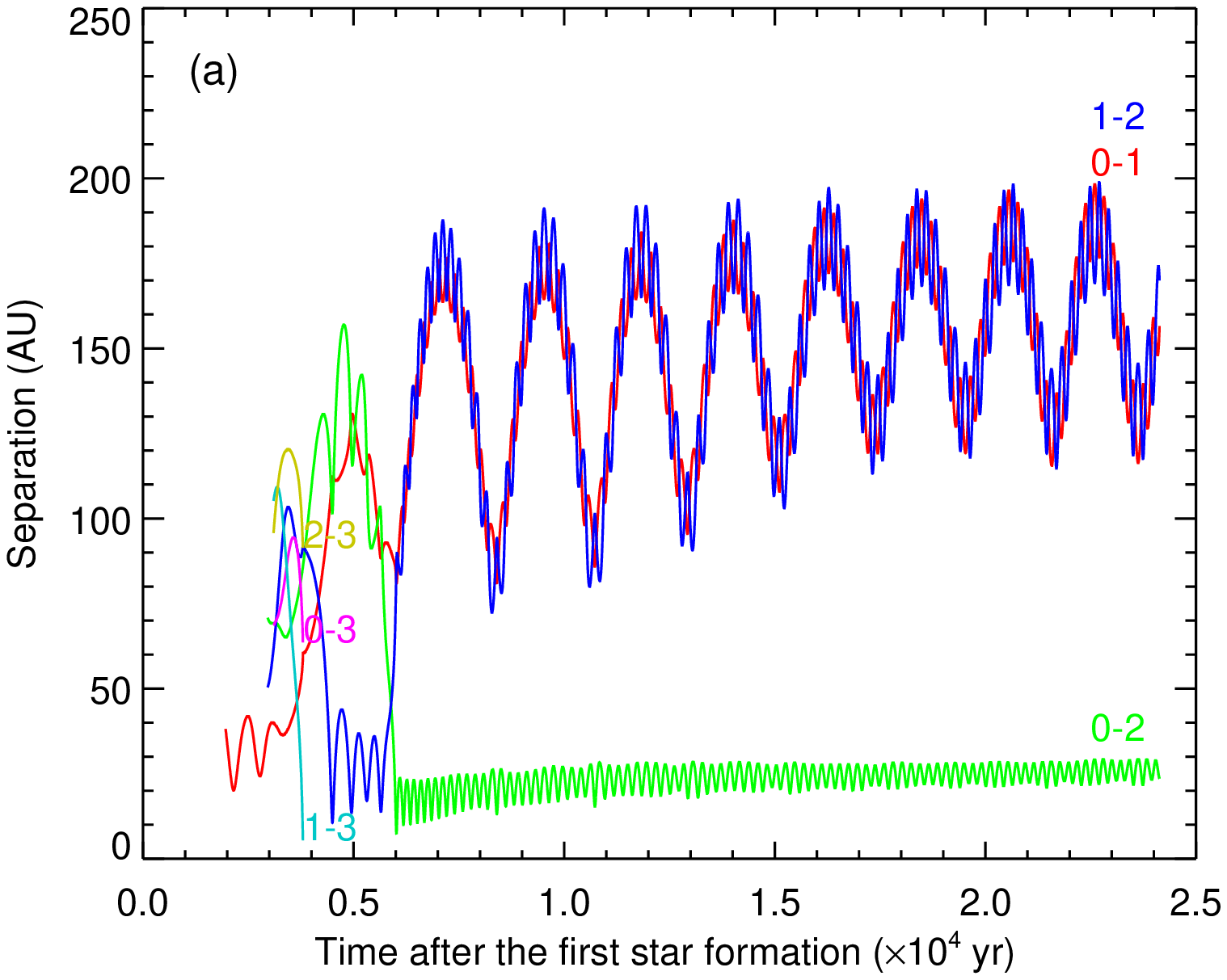}
\includegraphics[scale=0.5]{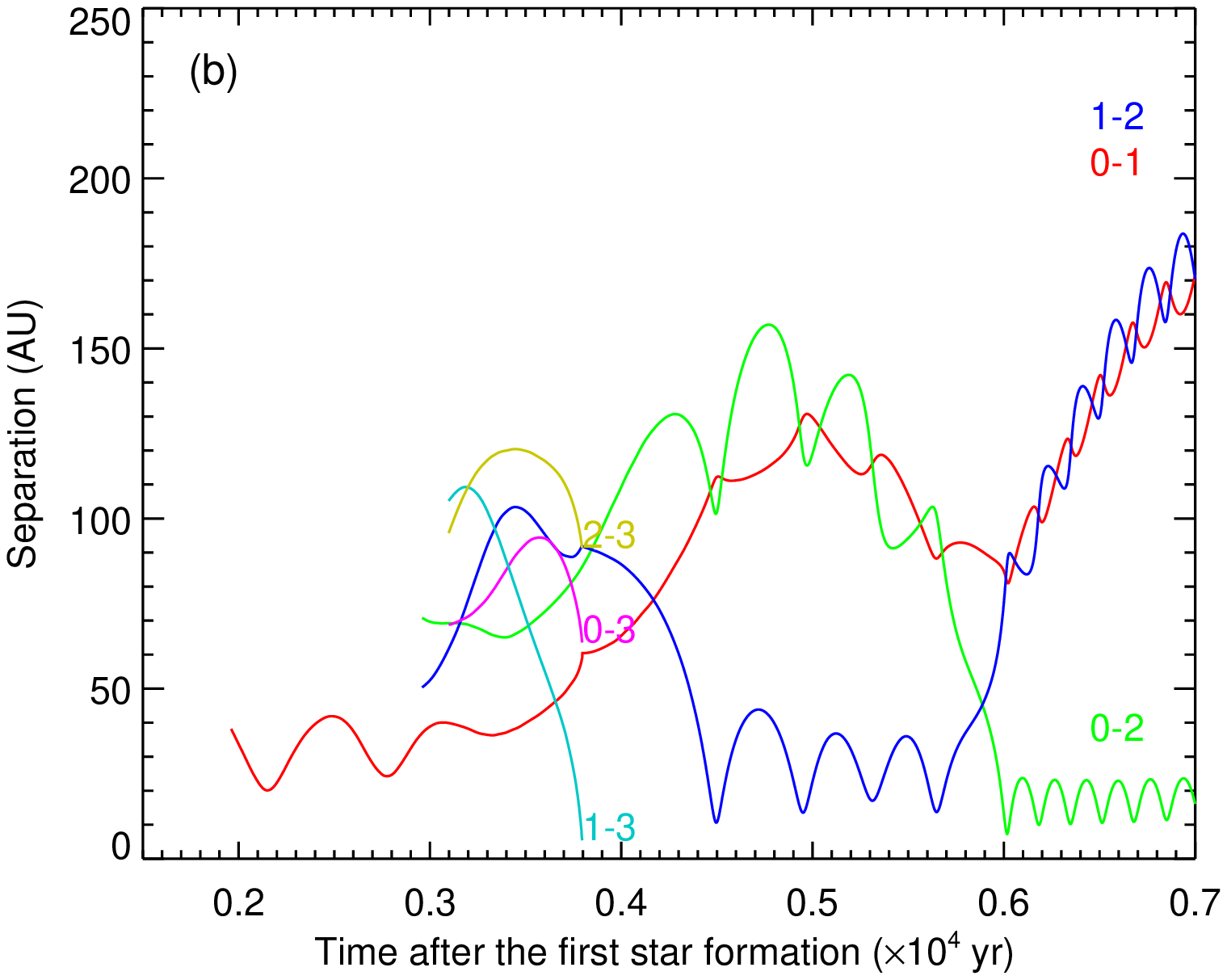}
\caption{ 
Separations between all of the pairs of sink particles as a function of time
after the first sink particle formation.
The label $n-m$ associated with each line
indicates the separation between Sinks~$n$ and $m$.
Panel ({\it b}) is an enlargement of the chaotic period in the early phase.
\label{time-pdist.eps}
}
\end{figure}

Figure~\ref{time-pdist.eps} shows the changes in separation between all
of the pairs of sink particles.  In the early stages, four sink particles were formed
and they exhibit chaotic orbits (Figure~\ref{time-pdist.eps}(b)).  In
their chaotic orbital motion, Sinks~1 and 3 merge at 
$t_p = 3800$~yr (the cyan line).
Subsequently, Sinks~1 and 2 form a close binary with
a separation of $10-40$~AU (the blue line).  At $t_p = 6000$~yr,
Sink~0 approached the close 
binary of Sinks~1 and 2.  Sink~0 and 1 were then exchanged to from a
new close binary with Sinks~0 and 2.  The orbit of Sink~1 then
gradually \revise{decreases the eccentricity from 0.3 to 0.2}
(the blue and red lines in Figure~\ref{time-pdist.eps}(a)).
The close binary of Sinks~0 and 2 has a high
eccentricity (the green line), leading to the misaligned circumstellar
disks.

\begin{figure}
\includegraphics[scale=0.6]{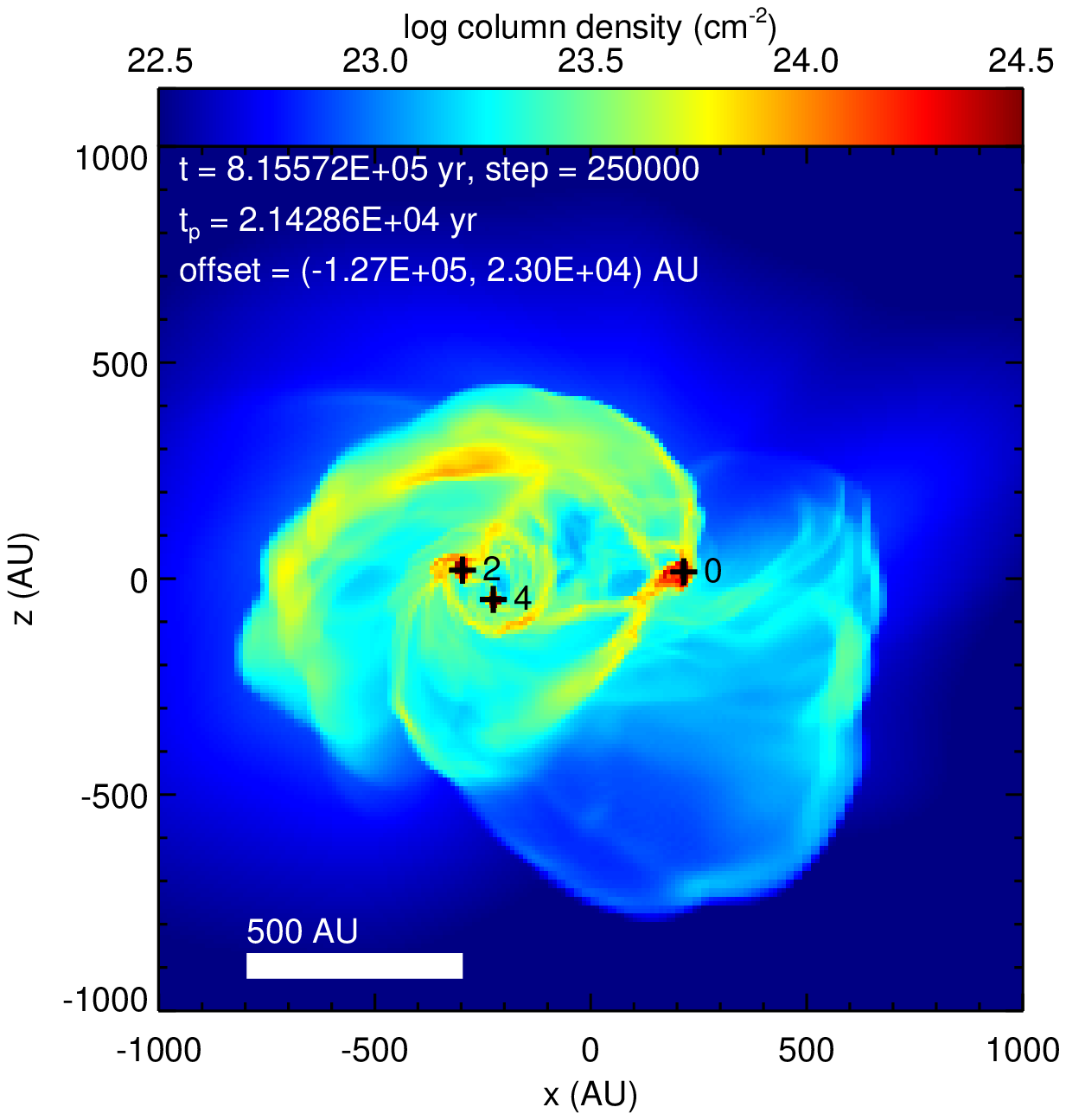}
\caption{ 
Column density distribution showing arc structures.
The black crosses indicate the positions of
the sink particles, which are labeled with identification numbers.
The realization is different from that shown in 
Figure~\ref{columndensity_plot.eps}.
\label{columndensity_plot_s2.eps}
}
\end{figure}

Because the sink particles have a chaotic epoch after
their formation, the density structure in the infalling envelope,
such as the arc, depends on the realization of the
initial turbulence.
We simulated the same model as the previous one but changed 
the random seed when generating the turbulence.
Figure~\ref{columndensity_plot_s2.eps} shows the column density
distribution at $t = 8.15 \times 10^5$~yr ($t_p = 2.14 \times 10^4$~yr).
By this stage, this realization model has produced six sink particles in total and they
eventually merge into three sink particles.
The sink particles undergo chaotic orbital motions until 
$t_p \simeq 2 \times 10^4$~yr, which is a considerably longer period than in the
previous model. 
This realization model also exhibits arc structures on a 1000~AU scale.
The line-of-sight velocities of the arc structures also range within
$\sim \pm 0.5\,\mathrm{km\,s}^{-1}$.
This model also exhibits misaligned circumstellar disks caused by 
close encounters between sink particles.  In the stage shown in 
Figure~\ref{columndensity_plot_s2.eps}, the most distant sink particle
(Sink 0) has a circumstellar disk inclined with respect to the other
circumstellar disks.

\section{Summary and Discussion}
\label{sec:discussion}

We demonstrated the formation of arc structure on a 1000~AU scale in the
infalling envelope during the early phase of star formation as a
consequence of the collapse of a weakly turbulent cloud core.
The density and velocity structures reproduced here are in agreement with
the arc structure revealed by the recent ALMA observations toward
MC27/L1521F. We also found that the misaligned disks are formed in
triple systems.

According to \citet{Tokuda14}, MC27/L1521F has three high-density
condensations observed by the dust continuum, one of which (MMS-1) was
also observed by the Spitzer space telescope \citep{Bourke06}.  The
other two condensations (MMS-2 and MMS-3) have high densities of
$10^{6-7}\,\mathrm{cm}^{-3}$, and MMS-2 is a candidate for the first
core as predicted by \citet{Larson69}.


A possible interpretation for these high-density condensations is that
they are formed by fragmentation during the protostellar collapse of
the cloud core as shown in our simulations.  The velocity of the arc
structure is ranged typically within $\sim \pm
0.5\,\mathrm{km\,s}^{-1}$, and it is consistent with the dynamical
velocity of the system, e.g., the Kepler velocity for a mass of
$0.2\,M_\sun$ and a radius of $500\,\mathrm{AU}$.  Moreover, the time
scale of the arc structure is
$1000~\mathrm{AU}/0.5~\mathrm{km\,s}^{-1} \simeq 10^4\,\mathrm{yr}$,
in agreement with the epoch shown in
Figure~\ref{columndensity_plot.eps}, $t_p = 1.760 \times
10^4$~yr.  However, a few issues remain unsolved in the following points.  
First,  
the masses of the sink particles 
grow exceeding $0.1\,M_\sun$ with this time scale.  These masses are
greater than those reported by \citet{Tokuda14}.  
Second, MMS-2, the
first core candidate, may have a short lifetime of less than 1000~yr
according to the recent radiation-transfer MHD simulations by
\citet{Tomida13}.  Within such a short period, an arc structure can
not extend to a 1000~AU scale with its typical velocity of $\sim
0.5\,\mathrm{km\,s}^{-1}$.


Protostars undergo chaotic orbits in the period of several
$\times 10^3 - 10^4$~yr after their formation.  Their chaotic orbits
influence the evolution of multiple stars as well as the surrounding
envelope, such as the arc structures.  The gravitational few-body
problem is therefore a key issue in the early phase of low-mass star
formation.  This is somewhat analogous to the interactions in ``the
competitive accretion scenario'' for massive star formation
\citep{Bonnell04}, in which the gravitational N-body problem is
supposed to govern the dynamics of clusters and therefore accretion.


Another scenario to consider for the formation of an arc structure
is a magnetic wall around the protostars.
The formation of a magnetic wall was
predicted by some theoretical works
\citep[e.g.,][]{Li96,Tassis05,Zhao11}.
The three-dimensional simulations showed that the magnetic
wall produces cavities in the infalling envelope.
The rims of the cavities have higher densities than the envelope 
and they may be observed as arc structures.
However, the magnetic field
suppresses fragmentation for the typical parameters
of molecular cloud cores \citep{Machida05}. The formation mechanism
of the high-density condensations, MMS-2 and MMS-3, can then not be
explained by the fragmentation process.
The observed arc structure therefore implies that the magnetic field
is weak enough to promote fragmentation of the cloud core.

Misalignment of a circumstellar disk is caused by the tidal
interaction with sink particles when they undergo close
encounters.  The close encounters are promoted by the chaotic orbits
in a triple system.  Thus, triple systems are expected to have
misaligned circumstellar disks, and to also be associated with arc
structures if they are 
young enough to be embedded in the dense envelope.

Misaligned disks were reported for
the young binary system, HK~Tau \citep{Jensen14},
AS~205 \citep{Salyk14},
and V2434Ori \citep{Williams14} based on the recent ALMA observations.
The young triple system, T~Tau, includes the tight binary of T~Tau~Sa
and T~Tau~Sb, and the existence of a misaligned circumstellar disk has
been indicated therein \citep{Skemer08,Ratzka09}.
Our model suggests that this misaligned disk could be caused by the tidal
interaction between the tight binary pair in the triple system.

\section*{Acknowledgments}

Numerical computations were carried out on Cray XC30 at the Center for Computational Astrophysics, National Astronomical Observatory of Japan.
This research was supported by
JSPS KAKENHI Grant Numbers
26400233,
26287030,
24244017,
23540270,
23403001,
23244027,
23103005,
22244014.

\label{lastpage}

\end{document}